\documentclass[11pt]{article}
\pdfoutput=1

\usepackage{euscript}
\usepackage{amssymb}
\usepackage{amsfonts}
\usepackage{amsbsy}
\usepackage{epsfig}
\usepackage{amsthm}
\usepackage{amscd}
\usepackage{amstext}
\usepackage{verbatim}
\usepackage{amsmath}
\usepackage{cancel}
\usepackage{capt-of}
\usepackage{empheq}
\usepackage{subfigure}

\usepackage[nosort]{cite}
\usepackage{bm}
\usepackage{authblk}
\usepackage{esint}
\usepackage[T1]{fontenc}

\usepackage[pdftex,linktoc=all]{hyperref}
\hypersetup{ 
colorlinks=true, 
linkcolor=black, 
citecolor=red, 
}

\def\ben{\begin{equation}}
\def\een{\end{equation}}

\let\a=\alpha \let\b=\beta   
   \let\k=\kappa

  \let\D=\Delta

\let\pa=\partial
\def\be{\begin{equation}}
\def\ee{\end{equation}}
\def\beq{\begin{equation}}
\def\eeq{\end{equation}}
\def\ba{\begin{array}}
\def\ea{\end{array}}

\def\dalemb#1#2{{\vbox{\hrule height .#2pt
       \hbox{\vrule width.#2pt height#1pt \kern#1pt
               \vrule width.#2pt}
       \hrule height.#2pt}}}

\newcommand{\bea}{\begin{eqnarray}}
\newcommand{\eea}{\end{eqnarray}}

\DeclareMathOperator{\sech}{sech}
\newcommand{\Tr}{{\rm Tr} }

\def\vep{{\varepsilon}}

\makeatletter
\newcommand*\bigcdot{\mathpalette\bigcdot@{.5}}
\newcommand*\bigcdot@[2]{\mathbin{\vcenter{\hbox{\scalebox{#2}{$\m@th#1\bullet$}}}}}
\makeatother

\renewcommand{\eqref}[1]{(\ref{#1})}

\def\ocal{{\mathcal{O}}}

\title{Wheeler-DeWitt states of the AdS-Schwarzschild interior}
\author{Sean A. Hartnoll}
\affil{
\it Department of Applied Mathematics and Theoretical Physics,  \\
\it University of Cambridge, Cambridge CB3 0WA, United Kingdom. \\
}
\date{}

\begin{document}

\maketitle

\begin{abstract}

We solve the Wheeler-DeWitt equation for the planar AdS-Schwarzschild interior in a minisuperspace approximation involving the volume and spatial anisotropy of the interior. A Gaussian wavepacket is constructed that is peaked on the classical interior solution. Simple observables are computed using this wavepacket, demonstrating the freedom to a choose a relational notion of `clock' in the interior and characterizing the approach to the spacelike singularity. The Wheeler-DeWitt equation may be extended out through the horizon, where it describes the holographic renormalization group flow of the black hole exterior. This amounts to the Hamilton-Jacobi evolution of the metric component $g_{tt}$ from positive interior values to negative exterior values. The interior Gaussian wavepacket is shown to evolve into the Lorentizan partition function of the boundary conformal field theory over a microcanonical energy window.

\end{abstract}

\newpage

\tableofcontents

\section{Introduction}

The boundary of an asymptotically AdS spacetime has a well-defined time coordinate $t$. This time coordinate can be extended into the bulk. However, events in the bulk are redshifted relative to the boundary by a factor of the bulk metric $g_{tt}$. This redshift typically increases monotonically with spatial distance from the asymptotic boundary \cite{Freedman:1999gp}. Thus, events that occur deeper in the bulk describe lower energy processes in the dual quantum field theory \cite{Susskind:1998dq}. The classical bulk equations of motion determine how bulk fields evolve with distance from the boundary, and this is called the holographic renormalization group \cite{Skenderis:2002wp}.

At a horizon $g_{tt} \to 0$, and therefore events at the horizon are infinitely redshifted with respect to the boundary. These describe the far infrared of the renormalization group. From the usual perspective of field theory renormalization, there is nothing left to integrate out and nowhere further to go. However, the bulk spacetime does not terminate at the horizon. It has recently been emphasized that in the bulk it is natural to continue the holographic renormalization group flow through the horizon into the black hole interior \cite{Frenkel:2020ysx}. In the interior this flow develops in time rather than space and extends all the way to the black hole singularity. The flow therefore can be thought of as a map from AdS boundary sources to scaling exponents near the interior cosmological singularity \cite{Frenkel:2020ysx}. This connection has been further elaborated in e.g.~\cite{Hartnoll:2020rwq, Hartnoll:2020fhc, Wang:2020nkd, Caceres:2021fuw, Mansoori:2021wxf, Bhattacharya:2021nqj, Das:2021vjf, Sword:2021pfm, Caceres:2022smh}.

One may hope that the AdS boundary perspective might shed light on the nature of the black hole interior and on the singularity in particular. This is not necessarily the same as asking how the interior is encoded in the boundary. The difficulty with encoding the interior in the boundary is, of course,
that the interior is causally disconnected from the boundary. Formally, this means that the interior time dependence corresponds to complex boundary energies that are accessed via analytic continuation \cite{Motl:2003cd, Fidkowski:2003nf, Hartman:2013qma, Grinberg:2020fdj}. A different perspective is to ask how the interior is prepared by the exterior. At a classical level this just amounts to solving the equations of motion with sources at the boundary \cite{Frenkel:2020ysx}. In this paper we will ask how the (semiclassical) quantum state of the interior is prepared by the AdS boundary.

The present paper is partially inspired by recent works \cite{Perry:2021mch, Perry:2021udd} that considered the Wheeler-DeWitt equation close to Kasner singularities of the kind that arise inside black holes. Historically, the vast majority of work on the Wheeler-DeWitt equation has been concerned with quantum cosmology and with fraught interpretational issues such as what it would mean to `predict' the state of the universe. See for example \cite{Halliwell:1989myn}. The interior of a black hole, in contrast, is prepared by its classical exterior. For a given exterior there should be, ultimately, no ambiguity regarding the state of the interior. This would seem to be, then, a promising setting in which to make sense of the Wheeler-DeWitt equation. The state of the interior may then be used to address important questions such as the quantum fate of classical spacelike singularities. 

`Eternal' black holes in asymptotically AdS spacetimes \cite{Maldacena:2001kr} provide an especially well-grounded starting point for exploration of the interior. Furthermore, planar AdS black holes offer a simple metric ansatz that explicitly incorporates both the asymptotically AdS scaling as well as near-singularity Kasner scaling \cite{Frenkel:2020ysx}. In this work we will discuss the Wheeler-DeWitt state of the planar AdS-Schwarzschild black hole interior within a mini-superspace approximation involving two metric functions: the scale factor and anisotropy of flat spatial slices at constant interior time.

\begin{figure}[h]
    \centering
    \includegraphics[width=0.9\textwidth]{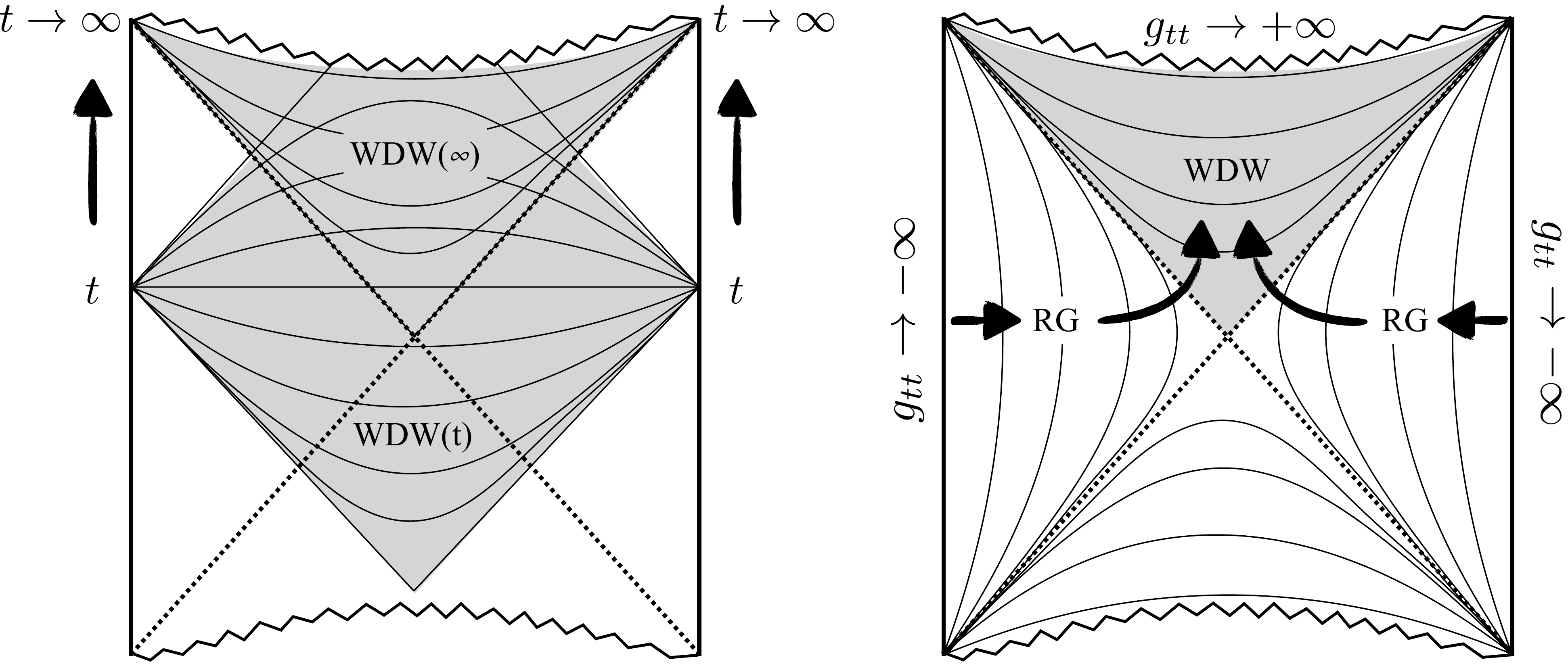}
    \caption{Two approaches to the black hole interior. Left: as boundary time $t$ becomes large, the `Wheeler-DeWitt patch', shown as a shaded region, coincides with the interior. Right:  Hamilton-Jacobi evolution along bulk radial slices can be parametrized, for example, by $g_{tt}$ which changes sign at the horizon. This evolution corresponds to holographic renormalization in the exterior and the Wheeler-DeWitt state in the interior. The arrows show the holographic renormalization preparing a state in the future interior, as opposed to the past. We discuss this ambiguity later.}
    \label{fig:penrose}
\end{figure}

Previous works have discussed the Wheeler-DeWitt equation in the context of asymptotically AdS black holes. Fig. \ref{fig:penrose} illustrates the relation between those works and what we will do. Important insights into black holes have been obtained by using the asymptotic boundary time to study the bulk as a conventional quantum mechanical system \cite{Hartman:2013qma, Maldacena:2013xja, Stanford:2014jda}. In particular, the bulk gravitational phase space can be obtained from the phase space of the boundary field theory \cite{Belin:2018fxe}. The boundary state at a given fixed boundary time $t$ is thereby associated to a region of the bulk called the
`Wheeler-DeWitt patch' in \cite{Brown:2015bva, Brown:2015lvg}.
From this perspective the state of the interior of the eternal black hole is obtained as the $t \to \infty$ limit of the thermofield double state of the boundary theory \cite{Belin:2018bpg}. See Fig.~\ref{fig:penrose}, left. This interior state should be the same as the state we will be discussing. As we have described in the first paragraphs above, the boundary `clock' has run out of time at this point. For this reason we view the interior state as `timeless' in the sense of being a cosmological solution to the Wheeler-DeWitt equation. To remain within a cosmological-like framework --- with Hamilton-Jacobi theory as the unifying mathematical description ---  we will relate the interior state to the holographic renormalization group flow of dual field theory, rather than its time dependence. See Fig.~\ref{fig:penrose}, right. These two perspectives are likely related, possibly through recently discussed $T^2$ deformations of the boundary theory \cite{Belin:2020oib, Araujo-Regado:2022gvw}.

This paper makes two technical points. The first is to solve the minisuperspace Wheeler-DeWitt equation in the interior. The general solution is found in \S\ref{sec:wdw}, and it is shown how classical interior physics is recovered by building a Gaussian wavepacket. The Wheeler-DeWitt wavefunction contains no time parameter but instead expresses correlations between geometric observables on spatial slices. We illustrate the use of several different choices of a relational `clock' with respect to which other observables may be computed: the volume of interior slices, their anisotropy, their extrinsic curvature, and their metric components $g_{tt}$ and $g_{xx}$. Using these clocks, we describe the approach to the singularity in \S\ref{sec:sing}.

These first results, regarding the behavior of the interior wavepacket, are independent of the exterior of the black hole. The second technical point we make, in \S\ref{sec:exterior}, is that the interior wavefunction is nothing other than the asymptotic (exterior) boundary Lorentzian partition function, extended to positive values of $g_{tt}$. This partition function is originally a function of negative $g_{tt}$, corresponding to the Lorentzian metric of the asymptotic boundary. In the minisuperspace we consider, however, the extension to positive $g_{tt}$ is simple; the basic quantity, the Hamilton-Jacobi function given in (\ref{eq:Sfull}) below, is linear in $g_{tt}$! One can think of the entire object as a single wavefunction labelled by a clock $g_{tt}$ that runs from $-\infty$ at the AdS boundary to $+\infty$ at the singularity, passing through zero at the horizon. This is illustrated in Fig.~\ref{fig:penrose}, right. The freedom to build wavepackets translates into different energy weightings of the boundary partition function. In particular, the Gaussian wavepacket corresponds to the partition function of a band of energies.

It is hoped that the framework in this paper can be used to provide both a solid conceptual grounding for quantum cosmological Wheeler-DeWitt equations and an approach towards understanding the quantum resolution of spacelike singularities in gravity. Future directions are discussed in \S\ref{sec:future}.

\section{AdS-Schwarzchild from the Hamilton-Jacobi equation}

We will be concerned with pure, four dimensional gravity with a negative cosmological constant. The action, including the Gibbons-Hawking boundary term, is
\be\label{eq:act}
S_\text{act} = \int d^4x \sqrt{-g} \left(R + 6 \right) + 2 \int d^3x \sqrt{h} K \,.
\ee
The subscript on $S_\text{act}$ is to differentiate the action from the Hamilton-Jacobi function that will be appearing shortly. Our first task will be to recover the well-known planar AdS-Schwarzschild solution to this theory from an approach that will generalize to semi-classical quantum states. We will restrict to metrics within the `minisuperspace' ansatz
\be\label{eq:ans}
ds^2 = - N^2 dr^2 + v^{2/3} \left( e^{4k/3} dt^2  + e^{-2k/3} \left[dx^2 + dy^2\right] \right) \,,
\ee
where $k,v,N$ are functions of $r$. We are interested, to start with, in the black hole interior, where $r$ is the time coordinate. The function $v$ is the spatial scale factor, which we will think of as the volume. The function $k$ determines the relative stretching of $g_{tt}$ relative to $g_{xx}$ ($= g_{yy}$ within our ansatz, retaining boundary rotational invariance), while keeping the spatial volume fixed. That is to say, it is a measure of anisotropy. This parametrization will turn out to be useful, as the Hamilton-Jacobi function and associated wave equations will be especially simple to work with. As we recall in \S\ref{sec:future}, generic black hole interiors are known to evolve into highly inhomogeneous spacetimes that are well beyond the anstaz (\ref{eq:ans}). Nonetheless, minisuperspace is a useful starting point from which to approach conceptual aspects of the problem.

On the ansatz (\ref{eq:ans}) the Lagrangian density becomes
\be\label{eq:Lag}
{\mathcal L} = 6 N v + \frac{2}{3} \frac{v^2 (\pa_r k)^{2} - (\pa_r v)^{2}}{N v} \,.
\ee
From this Lagrangian we may go to a Hamiltonian description. As expected, $N$ is a Lagrangian multiplier that imposes the Hamiltonian constraint:
\be\label{eq:Ham}
- \pi_k^2 + v^2 \pi_v^2 + 16 v^2 = 0 \,.
\ee
Here $\pi_k$ and $\pi_v$ are the conjugate momenta to $k$ and $v$.

The Hamilton-Jacobi equation is obtained by setting $\pi_k = \pa_k S$ and $\pi_v = \pa_v S$, so that
\be\label{eq:HJ}
- (\pa_k S)^2 + v^2 (\pa_v S)^2 + 16 v^2 = 0 \,.
\ee
It is sufficient to find a solution to this equation that has a single constant of integration, in addition to the constant giving an overall shift in $S$ (this shift trivially drops out of the equation, and we will not consider it).
One such solution is seen to be
\be\label{eq:Ssol}
S(v,k;k_o) = 4 v \sinh \left[k + k_o \right] \,, 
\ee
where $k_o$ is the desired constant of integration. As usual in Hamilton-Jacobi theory, the general solution to the equations of motion for the Lagrangian (\ref{eq:Lag}) is now obtained by setting the derivative of $S$ with respect to the constant of integration $k_o$ equal to a further constant, which we denote $\vep_o$:
\be\label{eq:diffS}
\pa_{k_o}S = \vep_o \,.
\ee
The solution to this last equation, using (\ref{eq:Ssol}), is
\be\label{eq:vsol}
v = \frac{\vep_o}{4} \sech \left[k + k_o \right] \,.
\ee

We may now verify that with (\ref{eq:vsol}) we have re-discovered the standard planar AdS-Schwarzschild black hole solution. To obtain the solution in a more recognizable form we may introduce a coordinate $z$ such that
\be
e^{2(k+k_o)} = \frac{\vep_o e^{k_o}}{2} z^3 - 1 \,.
\ee
Using this expression together with (\ref{eq:vsol}) and the equation of motion for $N$ that follows from (\ref{eq:Lag}), namely
$N^2 dr^2 = (v^2 dk^2 - dv^2)/(9 v^2)$, then the metric (\ref{eq:ans}) becomes
\be\label{eq:one}
ds^2 = \frac{1}{z^2} \left(- f(z) e^{-2 k_o} dt^2 + \frac{dz^2}{f(z)} + dx^2 + dy^2 \right)\,, \qquad  f(z) = 1 - \frac{\vep_o e^{k_o}}{2} z^3 \,.
\ee
Note that $f(z) < 0$ in the interior. The solution (\ref{eq:one}) is immediately recognized as the planar AdS-Schwarzschild geometry. This form of the solution is also valid in the black hole exterior, where $f(z)>0$ and $t$ becomes the timelike direction. In \S\ref{sec:exterior} below we will see that $\vep_o$ is the energy density of the black hole solution as defined at the AdS boundary.

In the interior we can note that, according to (\ref{eq:vsol}), the anisotropy $k$ goes from $-\infty$ at the horizon to $+\infty$ at the singularity. The spatial volume $v$ starts from $0$ at the horizon, expands to $\vep_o/4$ and then contracts back to $0$ at the singularity. The classical solution as given in (\ref{eq:vsol}) is relational, it
expresses one geometric property of time slices in terms of another, with no reference to any coordinate on the spacetime. This feature of Hamilton-Jacobi theory will be shared by the Wheeler-DeWitt wavefunction.

\section{Wheeler-DeWitt states of the interior}
\label{sec:wdw}

\subsection{Solutions and wavepackets}
\label{sec:packets}

We can now perform a semiclassical quantization of the minisuperspace introduced above. 
To start with we discuss the interior; the relationship to the exterior will be discussed in \S\ref{sec:exterior} below. The Wheeler-DeWitt state of the interior is obtained by promoting the momenta in the Hamiltonian constraint (\ref{eq:Ham}) to operators with the usual commutation relations. This gives a differential equation for the wavefunction $\Psi(v,k)$:
\be\label{eq:wdw}
\pa^2_k \Psi - v \pa_v \left(v \pa_v \Psi \right) + 16 v^2 \Psi = 0 \,.
\ee
There is a well-known ordering ambiguity in this equation.
The ordering does not affect the leading behavior in the classical limit, but does affect the leading quantum correction. We have adopted the prescription in which the differential operator that appears in the equation is the Laplacian (of the inverse DeWitt metric, see \S\ref{sec:metric} below) on superspace. A possible further term due to conformal coupling to the superspace scalar curvature vanishes in our case of a two dimensional minisuperspace \cite{Halliwell:1988wc, Halliwell:1989myn}. Note, furthermore, that the constant $r$ slices in (\ref{eq:ans}) are flat, with vanishing curvature.

Equation (\ref{eq:wdw}) can be solved by Fourier transforming in $k$.
This is a first instance where the 
choice of $k,v$ coordinates is helpful, as they have separated the equation.
The Fourier modes, with label $\vep$, are given by modified Bessel functions $e^{i \vep k} I_{\pm i \vep}(4 v)$. The same Bessel function solutions have recently arisen in the Wheeler-DeWitt equation for the flat space Schwarzschild interior \cite{Bouhmadi-Lopez:2019kkt, Yeom:2021bpd} and also in Jackiw-Teitelboim gravity \cite{Harlow:2018tqv, Yang:2018gdb}. In these latter papers the slices of the bulk geometry extend to the exterior boundary and thereby inherit a preferred time. In our minisuperspace context, the Wheeler-DeWitt equation should not be taken seriously beyond the semiclassical regime and it will be more useful to to consider the semiclassical limit of these solutions. These are obtained using conventional one dimensional WKB methods as
\be\label{eq:Psi}
\Psi_\pm(v,k) = \int \frac{d\vep}{2 \pi} \a_\pm(\vep) \psi_\pm(v,k;\vep) \,.
\ee
The functions $\alpha_\pm(\vep)$ are arbitrary and, in the oscillating regime $4 v < \vep$, the modes
\be\label{eq:psiep}
\psi_\pm(v,k;\vep) = \frac{e^{i \vep k}}{(\vep^2 - 16 v^2)^{1/4}} \exp\left\{
\pm i \left[ \sqrt{\vep^2 - 16 v^2} - \vep \tanh^{-1} \frac{\sqrt{\vep^2 - 16 v^2}}{\vep} - \frac{\pi}{4} \right] \right\} \,.
\ee
To connect with standard WKB formulae note that the term in square brackets in (\ref{eq:psiep}) is $\int \frac{dv}{v} \sqrt{\vep^2 - 16 v^2}$. It is natural here to introduce $V = \log v$, so that the measure $dv/v = dV$. From this point of view, each $\vep$ mode describes a particle with energy $\vep^2$ in an exponential potential $16 e^{2 V}$. Scattering in an exponential potential is familiar from Liouville theory \cite{Ginsparg:1993is}, where the modified Bessel functions mentioned above are again the exact solutions. A given $\psi_\pm(v,k;\vep)$ state has oscillatory behavior over the range of volumes $v \leq \vep/4$ that occur in the classical interior of a black hole with asymptotic AdS energy density $\vep$.

The modes (\ref{eq:psiep}) are completely delocalized in the $k$ direction. This state, therefore, does not capture the classical evolution (\ref{eq:vsol}) of the interior towards the singularity as $k \to \infty$. That is to say, the classical evolution is localized in both $k$ and $v$, and indeed also in the conjugate momenta $\pi_k$ and $\pi_v$, while these states are not. To construct such a classical state we need to form wavepackets out of the $\psi_\pm(v,k;\vep)$ modes, by choosing appropriate functions $\alpha_\pm(\vep)$ in (\ref{eq:Psi}). This is familiar from standard quantum mechanics, and has also been widely discussed in quantum cosmology, see e.g.\cite{Hartle:1989zcl} and references therein. The natural wavepacket to build is a Gaussian superposition of WKB modes peaked around some energy $\vep_o$ with width $\Delta$:
\be\label{eq:D2}
\Psi_{\pm, \Delta,\vep_o,k_o}(v,k) =  c \int \frac{d\vep}{2\pi} e^{i k_o \vep} e^{- (\vep - \vep_o)^2/(2\Delta^2)} \psi_\pm(v,k;\vep) \,.
\ee
Here the normalization constant $c^2 = 2 \sqrt{\pi}/\Delta$. Norms will be discussed shortly.

To start to uncover the physics of the state (\ref{eq:D2}), it is helpful to connect to the classical discussion in the previous section. To do this most explicitly, we can represent (\ref{eq:D2}) in a different way. We first note that, in generality, the states (\ref{eq:Psi}) can be expressed in a Fourier transformed basis. Fourier transforming $\alpha_\pm(\vep)$ in (\ref{eq:Psi}) and then performing the $\vep$ integral by stationary phase leads to
\be\label{eq:Psi2}
\Psi_\pm(v,k) = \int \frac{d \bar k}{2\pi} \beta_\pm(\bar k) \phi_\pm(v,k;\bar k) \,.
\ee
Here $\b_\pm$ is the inverse Fourier transform of $\a_\pm$ and the modes
\be\label{eq:psik}
\phi_\pm(v,k;\bar k) = e^{\pm i S(v,k;\bar k)} \,,
\ee
where $S(v,k;\bar k) = 4 v \sinh(k + \bar k)$ is the solution to the classical Hamilton-Jacobi equation found previously in (\ref{eq:Ssol}). The quadratic fluctuations of $\vep$ about the stationary phase point cancel the functional dependence on $v$ in the prefactor of the WKB modes (\ref{eq:psiep}). There is correspondingly no nontrivial prefactor in (\ref{eq:psik}). In fact, (\ref{eq:psik}) are exact solutions to the Wheeler-DeWitt equation (\ref{eq:wdw}). The representation (\ref{eq:Psi2}) of the general solution in terms of elementary functions is especially easy to work with. The exact modes (\ref{eq:Psi2}) are related to the Bessel function modes mentioned previously by Fourier transformation. For example
\be\label{eq:K}
\int_{-\infty}^\infty e^{4 i v \sinh k} e^{-i \vep k}dk = \frac{i \pi e^{\pi \vep/2}}{\sinh (\pi\vep)} \left[I_{i\vep}(4v) - I_{-i\vep}(4v) \right] = 2 e^{\pi \vep/2} K_{i \vep}(4v)\,.
\ee
It is noteworthy here that the $+$ sign mode in the decomposition (\ref{eq:Psi2}) Fourier transforms into a sum of $+$ and $-$ sign modes in the decomposition (\ref{eq:Psi}). The sum of modes in (\ref{eq:K}) is natural in conventional quantum mechanics, describing an incoming beam that reflects off the potential in (\ref{eq:wdw}). However, below we will see that this is not necessarily the case in the Wheeler-DeWitt equation due to the need to obtain positive probabilities from a second order equation. The choice of modes will depend upon the choice of `clock'.

For the Gaussian wavepacket (\ref{eq:D2}) we have that the function in (\ref{eq:Psi2}) is
\be\label{eq:betag}
\beta_{\pm, \Delta,\vep_o,k_o}(\bar k) = \bar c \,  e^{-i \vep_o (\bar k - k_o)} e^{- (\bar k - k_o)^2 \Delta^2/2} \,.
\ee
Here $\bar c^{2} = 2 \pi^{1/2} \Delta$.
Using this form in (\ref{eq:Psi2}) describes a Gaussian spread in $\bar k$ of width $1/\Delta$ about the value $k_o$. If this spread is small, then the $\bar k$ integral in (\ref{eq:Psi2}) is seen to be dominated by the stationary phase point $\pm \pa_{\bar k}S = \vep_o$. This is precisely the classical solution as given in (\ref{eq:diffS}) and discussed at length in the previous section (we will comment shortly on the meaning of the two different signs). Thus we find that, as expected, the wavepacket describes a solution that is strongly supported on the classical interior evolution. This fact will be established in detail in the following few subsections, where we will also characterize the quantum variance about the classical solution introduced by the Gaussian smearing. As is usual in quantum mechanics, if the wavepacket is too narrow in any given representation, the variance of the conjugate momentum will become large. To keep the quantum variance of all simple observables small compared to their classical expectation values requires both the wavepacket (\ref{eq:betag}) and its Fourier transform (\ref{eq:D2}) to be narrow. Specifically, we require
\be\label{eq:range}
\vep_o \gg \Delta \gg 1 \,.
\ee
This is only possible in the semiclassical regime $\vep_o \gg 1$. The value of $k_o$ is not constrained.

\subsection{Clocks and measures}

Before discussing the wavefunction (\ref{eq:D2}) further, we should say a few words about the choice of $\pm$ sign. Flipping the sign in the exponent of the wavefunction flips the expectation values of momentum operators, that act as derivatives. The two different signs are therefore naturally associated to the oppositely evolving future and past black hole interiors, respectively. In a quantum cosmological context this is commonly discussed in terms of an expanding and a contracting universe. However, as we will recall shortly, using the canonical conserved inner product given by the Wheeler-DeWitt equation only one set of these modes, with a fixed sign, admits a positive definite norm. This suggests that only one of the interiors may be described at a time within a conventional probabilistic interpretation of quantum mechanics. In the semiclassical regime this fact is not of practical importance because superpositions of two states moving in different directions will decohere into distinct branches of the wavefunction. Decoherence has been widely discussed in the context of quantum cosmology, e.g. \cite{PhysRevD.39.2912, PhysRevD.39.2924}. Equivalently, a wavefunction with support on both branches may be said to exhibit spontaneous breaking of time reversal invariance \cite{BANKS1985332, Araujo-Regado:2022gvw}. All told, to discuss the physics as the singularity is approached in the future interior, for example, it is physically sensible to focus on a state with a given sign, such as the wavepacket $\Psi_{+, \Delta,\vep_o,k_o}(v,k)$.

We are considering the Wheeler-DeWitt equation as a `timeless' description of the black hole interior, as discussed in the introduction above. A choice of slicing of the interior, however, gives a relational notion of time. That is to say, one picks a given coordinate on superspace to act as the clock and one interprets the Wheeler-DeWitt wavefunction as giving probabilities conditioned on the value of the clock. For this framework to make sense it is essential that probabilities are conserved under the relational time evolution. The Wheeler-DeWitt equation ensures this \cite{PhysRev.160.1113}, as we now describe.

To start with, we consider the geometric quantities $k$ and $v$ as possible clocks. These turn out to be technically convenient for evaluating expectation values. We will go on to consider several further possible clocks, each of which brings its own advantages and subtleties. If we choose $k$ as the clock then the conserved norm is
\be\label{eq:pk}
|\Psi|_k^2 = \frac{-i}{2} \int \frac{dv}{v} \left(\Psi^* \pa_k \Psi - \Psi \pa_k \Psi^* \right) \,.
\ee
Using the Wheeler-DeWitt equation (\ref{eq:wdw}) it is immediately seen that $\pa_k(|\Psi|_k^2) = 0$,
so long as the wavefunction decays at large and small $v$ with $k$ fixed. This establishes that the norm is conserved under evolution in $k$. The volume $v$ is constrained to be positive. This makes it natural, as mentioned previously, to work in terms of the unconstrained coordinate $V = \log v$. However, it will be technically more convenient to take $v$ to run from $-\infty$ to $+\infty$ in (\ref{eq:pk}), even though physical volumes are positive. This is because many integrals then become simple Fourier transforms. The extension of the integral to $v < 0$ is harmless so long as quantities are subsequently computed in states that vanish at negative $v$. The Gaussian wavepackets that we will be considering have exponentially small tails at negative $v$, and therefore the error made by integrating over negative $v$ is negligible in the semiclassical limit. The norm (\ref{eq:pk}) is then most easily evaluated on a general state in the representation (\ref{eq:Psi2}), to obtain
\be\label{eq:knorm}
|\Psi_\pm|_k^2 = \pm \int \frac{d\bar k}{2\pi} |\beta_\pm(\bar k)|^2 \,.
\ee
As advertised above, this norm is positive definite and gives a good notion of probability on the states with a $+$ sign. In particular, $|\beta_+(\bar k)|^2$ can be interpreted as the probability that the interior universe is following a classical trajectory with anisotropy offset $\bar k$. The norm (\ref{eq:knorm}) is explicitly independent of $k$ and therefore manifestly conserved. 

If we instead chose $v$ as the clock then the conserved norm is now
\be\label{eq:vnorm}
|\Psi|_v^2 = \frac{-i}{2} \int dk \left(\Psi^* v \pa_v \Psi - \Psi v \pa_v \Psi^* \right) \,.
\ee
Again using the Wheeler-DeWitt equation (\ref{eq:wdw}) we have that $\pa_v(|\Psi|_v^2) = 0$, assuming that the wavefunction decays at large $|k|$ with $v$ fixed. The norm is therefore conserved under evolution in $v$.
With this norm it is more convenient to use the 
decomposition (\ref{eq:Psi}) of a general state, leading to
\be\label{eq:nplus}
|\Psi_\pm|_v^2 = \pm \int \frac{d\vep}{2\pi} |\alpha_\pm(\vep)|^2 \,.
\ee
As previously, this norm is positive definite and gives a good notion of probability on the states with a $+$ sign. Thus $|\alpha_+(\vep)|^2$ can be interpreted as the probability that the interior universe is following a classical trajectory with `energy' $\vep$.

There is a subtle and important point in going from (\ref{eq:vnorm}) to (\ref{eq:nplus}). As we noted above, the classical interior solution reaches a maximal volume of $v = 4 \vep$. The volume is not single-valued in the interior, vanishing both towards the horizon and towards the singularity. In standard quantum mechanics, prior to building wavepackets, the WKB version of this evolution would be an incoming wave that bounces off the potential barrier at $v = 4 \vep$, leading to a reflected wave going in the opposite direction. The resulting wavefunction would decay exponentially in the classical disallowed region $v > 4 \vep$.
This is indeed precisely what happens with the modes (\ref{eq:Psi2}) that we used above in obtaining the norm (\ref{eq:knorm}) for the $k$ clock. As we saw in 
(\ref{eq:K}), the positive probability modes under that norm contain waves going in both directions and decay in the classically disallowed region. So far so good. However, things work out in a less familiar way in the $v$ clock, as we now explain.

The norm (\ref{eq:vnorm}) for the $v$ clock is the flux in the $v$ direction. The subspace of states on which this is positive have an incoming wave with no reflected wave.
Thus, reflection is not allowed under this norm. We wrote down the oscillating modes in (\ref{eq:psiep}). By the usual WKB matching procedure, these modes will necessarily contain a (complex) sum of exponentially growing and decaying terms in the disallowed region. While this behavior would not be normalizable under the usual quantum mechanical norm, these states are normalizable with respect to (\ref{eq:vnorm}): the exponentially growing and decaying terms multiply each other in such a way that the classically disallowed region gives the same contribution to (\ref{eq:nplus}) as the oscillating region. Thus, in particular, the integral over $\vep$ in (\ref{eq:nplus}) runs over all values, unconstrained by $v$. Consequently, the norm (\ref{eq:nplus}) is explicitly independent of $v$ and therefore manifestly conserved. One may be concerned that the classically disallowed region contributes to the norm. However,
we will see below that semiclassical expectation values are nonetheless dominated by contributions from the classical allowed region.

Now, in fact, because $\alpha_+$ and $\beta_+$ are Fourier transforms of each other we have
\be
|\Psi_+|_v^2 = |\Psi_+|_k^2 \,.
\ee
The two clocks lead to the same norm. Expectation values will not be identical, however. This is clear because with one choice of clock expectation values are computed given a precise value of $k$ while for the other clock a precise value of $v$ is given.

In the following few sections we illustrate the computation of expectation values with several simple examples. One objective here is to gain some confidence in the consistency of the answers obtained using various different clocks. We will also comment on a few subtle points as we go along. The first step is to express expectation values involving $k,v,\pi_k,\pi_v$ in terms of the $\a$ and $\beta$ coefficients. As we have just seen, these coefficients determine the probability of finding the interior in different classical states. The expectation values we will compute can be expressed as functions averaged over these probabilities. In the remainder we only consider states with the $+$ sign and hence will drop the $+$ label.

\subsection{Expectation values for the anisotropy clock}
\label{sec:aniso}

In this section $k$ is taken as the clock. We may now use the $k$-norm (\ref{eq:pk}) to calculate the expectation values of observables conditioned upon the given value of $k$.
For example, the expectation value of the volume is evaluated to be
\be\label{eq:vk}
\langle v \rangle_k = \frac{i}{2} \int \frac{d\bar k}{2\pi} \frac{\b^*(\bar k) \beta'(\bar k) - \b^{*\prime}(\bar k) \b(\bar k)}{4 \cosh(k + \bar k)} \,.
\ee
Using the Gaussian wavepacket (\ref{eq:betag}) we can expect that the integral will be dominated by values of $\bar k$ within a range $1/\Delta$ of $k_o$. Expanding in $\Delta \gg 1$ we indeed obtain
\be\label{eq:vvev}
\langle v \rangle_{k;\Delta,\vep_o,k_o} = \frac{\vep_o}{4} \sech(k + k_o) + \ocal\left(1/\Delta^2 \right) \,.
\ee
The leading term here is precisely the classical solution (\ref{eq:vsol}). When $\Delta \gg 1$ the expectation value is strongly localized on the classical solution and corrections to the volume are small, uniformly in $k$. All $1/\Delta$ corrections below are also uniformly bounded in $k$. However, there is also a quantum variance to the volume. This is obtained from
\be\label{eq:vv2}
\langle v^2 \rangle_k =  \int \frac{d\bar k}{2\pi} \frac{|\beta'(\bar k)|^2 - \frac{1}{4} \pa^2_{\bar k}(|\beta(\bar k)|^2)}{16 \cosh^2(k+\bar k)} \,.
\ee
As we explained around (\ref{eq:range}) above, keeping the variance small on the Gaussian wavepacket requires $\vep_o \gg \Delta \gg 1$. Evaluating (\ref{eq:vv2}) in this regime we obtain the variance
\be
\left. \frac{\text{var}(v)_k}{\langle v \rangle_{k}^2} \right|_{\Delta,\vep_o,k_o} = \left. \frac{\langle v^2 \rangle_{k} - \langle v \rangle_{k}^2}{\langle v \rangle_{k}^2} \right|_{\Delta,\vep_o,k_o}= \frac{\Delta^2}{2 \vep_o^2} + \ocal\left(1/\Delta^2\right) \,.
\ee
The variance is seen to be uniformly small compared to the expectation value in this regime.

We can also look at the momentum conjugate to the volume: $\pi_v = - i \pa_v$. There is an ordering ambiguity with the computation of the expectation value. The most natural prescription, that gives a real answer, is
\begin{align}
\langle \pi_v \rangle_k & = \frac{-1}{2} \int dv \left(\Psi^* \pa_v \left[\frac{\pa_k \Psi}{v}\right] - \left[\pa_v\Psi\right] \frac{\pa_k \Psi^*}{v}  \right) \\
& = \int \frac{d\bar k}{2\pi} |\beta(\bar k)|^2 4 \sinh(k + \bar k) \,.
\end{align}
Using the wavepacket (\ref{eq:betag}), we obtain
\be
\langle \pi_v \rangle_{k;\Delta,\vep_o,k_o} = 4 \sinh(k + k_o) + \ocal\left(1/\Delta^2 \right) \,.
\ee
The leading term is again the classical solution: from (\ref{eq:Ssol}) $\pi_v = \pa S/\pa v = 4 \sinh[k+k_o]$. For $\Delta \gg 1$ the corrections are again uniformly small. The variance is obtained using
\be
\langle \pi_v^2 \rangle_k = \int \frac{d\bar k}{2\pi} |\beta(\bar k)|^2 16 \sinh^2(k + \bar k) \,.
\ee
On the Gaussian wavepacket it is found that $(\langle \pi_v^2 \rangle_k - \langle \pi_v \rangle_k^2)/\langle \pi_v \rangle_k^2 \sim 1/\Delta^2$, so that $\Delta \gg 1$ is a sufficient condition for this momentum to be classical.

Using the $k$ clock, the value of $k$ itself is known precisely by assumption. The momentum $\pi_k = - i \pa_k$ generates `time translations' for this clock and is therefore analogous to the Hamiltonian. The expectation value
\begin{align}
\langle \pi_k \rangle_k & = \frac{-1}{2} \int \frac{dv}{v} \left(\Psi^* \pa_k^2 \Psi - \pa_k \Psi^* \pa_k \Psi \right) \\
& = \frac{1}{2} \int \frac{dv}{v} \left(|\pa_k \Psi|^2 + v^2 |\pa_v \Psi|^2 + 16 v^2 |\Psi|^2 \right) \\
& = \frac{i}{2} \int \frac{d\bar k}{2\pi} \left[\b^*(\bar k) \beta'(\bar k) - \b^{*\prime}(\bar k) \b(\bar k) \right] \,. \label{eq:pikfin}
\end{align}
To obtain the second line we used the Wheeler-DeWitt equation (\ref{eq:wdw}).
For the Gaussian wavepacket we obtain
\be\label{eq:pikk}
\langle \pi_k \rangle_{k;\Delta,\vep_o,k_o} = \vep_o \,.
\ee
There are no $1/\Delta$ corrections to this expression. Thus we see that $\vep_o$ has a double life as an energy. It is the energy of the asymptotically AdS black hole and is also the energy of the interior Wheeler-DeWitt state with respect to time evolution by $k$. We will see later that the constant $k_o$, conjugate to $\vep_o$ in (\ref{eq:diffS}), correspondingly has a double life as a boundary as well as an interior time. The expectation value (\ref{eq:pikk}) of the energy is time-independent, as it should be because the Wheeler-DeWitt equation (\ref{eq:wdw}) is invariant under shifts in $k$. Using (\ref{eq:pikk}) together with (\ref{eq:vvev}) gives, to leading order at large $\Delta$ on the Gaussian wavepacket, $\langle \pi_k \rangle_k = 4 \langle v \rangle_k \cosh(k + k_o)$. This is in agreement with the classical relation following from (\ref{eq:Ssol}) and $\pi_k = \pa_k S$. The variance of the energy is also time-independent, as can be seen from
\be
\langle \pi_k^2 \rangle_k =  \int \frac{d\bar k}{2\pi} |\beta'(\bar k)|^2 \,.
\ee
For the Gaussian wavepacket this gives $\langle \pi_k^2 \rangle_k = \langle \pi_k \rangle_k^2 + \frac{1}{2} \Delta^2$. This quantum variance is small when $\vep_o \gg \Delta$.

\subsection{Expectation values for the volume clock}

We may now consider $k$ and $\pi_k$ using the $v$ clock. For the momentum one finds
\be
\langle \pi_k \rangle_v = \int \frac{d\vep}{2\pi} |\a(\vep)|^2 \vep \,.  
\ee
This is manifestly equal to the expression (\ref{eq:pikfin}) computed in the other norm, using the fact that $\a$ and $\beta$ are related by a Fourier transform. Similarly $\langle \pi_k^2 \rangle_v = \langle \pi_k^2 \rangle_k$.
This momentum is the same, and conserved, in both norms.

For $k$ itself we find, working within the semiclassical WKB approximation,
\be\label{eq:kv}
\langle k \rangle_v = \int \frac{d\vep}{2\pi} \left(\frac{i}{2} \left[\a^*(\vep) \a'(\vep) - \a^{*\prime}(\vep) \a(\vep)\right] + |\a(\vep)|^2 \tanh^{-1} \frac{\sqrt{\vep^2 - 16 v^2}}{\vep}\right) \,.
\ee
Evaluated on the Gaussian wavepacket, now using (\ref{eq:D2}) and expanding in $\Delta/\vep_o \ll 1$,
\be\label{eq:kvd}
\langle k \rangle_{v;\Delta,\vep_o,k_o} = - k_o + \sech^{-1}\frac{4v}{\vep_o} - \frac{\Delta^2}{\vep_o^2} \frac{1}{4 [1 - (4 v/\vep_o)^2]^{3/2}} + \cdots \,.
\ee
The first two terms on the right hand side reproduce the classical solution (\ref{eq:vsol}). Here we re-expressed $\tanh^{-1}$ in terms of $\sech^{-1}$. The expansion in $\Delta/\vep_o$ breaks down close to the turning point at $v = \vep_o/4$, which is the maximal volume slice. This occurs because, as usual, the WKB approximation breaks down at turning points. A sensible answer can be obtained using the full modes written in terms of modified Bessel functions. Deep in the non-classical regime of small $v$ one can again use the WKB expression continued to the exponential rather than oscillatory region. This amounts to asking for the expectation value of $k$ conditioned upon the value of $v$ being classically unallowed. This is not an especially natural question to ask, but recall from the discussion below (\ref{eq:nplus}) that the exponential region does contribute to the norm of the state. However, perhaps satisfyingly, one finds that the second term in (\ref{eq:kv}) vanishes to leading WKB order in this regime, so that $\langle k \rangle_v = - k_o$ for $v > \vep_o/4$. There is no volume dependence in the classically forbidden regime, consistent with the simplest continuation of (\ref{eq:kvd}) to small $v$, in which $\langle k \rangle$ gets stuck at $-k_o$.

The value of $v$ itself is specified as the `time'. The momentum $\pi_v$ is the `Hamiltonian' for the $v$ clock. Because the Wheeler-DeWitt equation depends explicitly on $v$ we may expect this Hamiltonian to be time dependent. The expectation value
\begin{align}
\langle \pi_v \rangle_v & = \frac{-1}{2} \int dk \left(\Psi^* \pa_v ( v \pa_v \Psi)) - \pa_v \Psi v \pa_v \Psi^* \right) \\
& = \frac{1}{2} \int \frac{dk}{v} \left(|\pa_k \Psi|^2 + v^2 |\pa_v \Psi|^2 - 16 v^2 |\Psi|^2 \right) \\
& = \int \frac{d\vep}{2\pi} |\a(\vep)|^2 \sqrt{\vep^2/v^2 - 16} \,.
\end{align}
On the Gaussian wavepacket
\be\label{eq:pivv}
\langle \pi_v \rangle_{v;\Delta,\vep_o,k_o} = \sqrt{\vep_o^2/v^2 - 16}
- \frac{\Delta^2}{\vep_o^2} \frac{4 v}{[1 - (4 v/\vep_o)^2]^{3/2}} + \cdots \,.
\ee
The leading behavior is consistent with the classical solution, using $\pi_v = 4 \sinh(k + k_o)$ as well as the solution (\ref{eq:vsol}). The expansion in $\Delta/\vep_o$ again breaks down close to the turning point, similarly to the discussion in the previous paragraph. Again using the exponential form of the wavefunctions beyond the turning point, the Hamiltonian is seen to vanish in the classically disallowed region -- extending (\ref{eq:pivv}) continuously --  which is perhaps intuitive.

\subsection{York time}
\label{sec:york}

As we have seen, the spatial slice volume $v$ is an awkward quantity to work with. It should be constrained to be positive and is not single-valued in the classical interior. It has long been recognized that a nicer choice of `time' on superspace is the trace $K$ of the extrinsic curvature of the spatial slices \cite{PhysRevLett.28.1082}. The trace of the extrinsic curvature is proportional to the momentum conjugate to volume, and we will work in terms of
\be\label{eq:kap}
\kappa \equiv \frac{\pi_v}{4} \,.
\ee
The factor of $4$ leads to cleaner expressions below. We will see that the wavepackets we are considering become especially simple when represented as a function of $\kappa$.

We may directly write down the Wheeler-DeWitt equation in volume `momentum space'. Recall that the Wheeler-DeWitt equation (\ref{eq:wdw}) arose from quantising the Hamiltonian constraint in a position basis for a wavefunction $\Psi(v,k)$. However, we could also have written down this equation in a momentum basis for the volume, leading to an equation for $\hat \Psi(\kappa,k)$:
\be\label{eq:momwdw}
\pa_k^2 \hat \Psi - \pa_\k \left[\k \pa_\k \left(\k \hat \Psi \right) \right] - \pa_\k^2 \hat \Psi = 0 \,.
\ee
This equation is equivalent to
\be\label{eq:wdw3}
\pa_k^2 \hat \Psi - \frac{1}{\sqrt{1 + \k^2}} \frac{d^2}{d (\sinh^{-1}\k)^2} \left(\sqrt{1 + \k^2} \hat \Psi\right) = 0 \,,
\ee
which is just a one-dimensional wave equation. We can immediately write down the general solution in terms of left- and right-moving modes:
\be\label{eq:Fgen}
\hat \Psi_\pm(\kappa,k) = \frac{F_\pm\left(k \mp \sinh^{-1}\kappa \right)}{\sqrt{1 + \k^2}}  \,.
\ee
Here the $F_\pm$ are any function.

We may alternatively obtain the general solution (\ref{eq:Fgen}) by Fourier transforming the `position space' modes (\ref{eq:psik}):
\be\label{eq:phikappa}
\hat \phi_\pm(\kappa,k; k_o) = \frac{2}{\pi} \int dv e^{-i 4 \kappa v} \phi_\pm(v,k; k_o) =  \delta\left(\kappa \mp \sinh[k + k_o]\right) \,.
\ee
As remarked upon above, we have integrated the volume $v$ from $-\infty$ to $+\infty$. Using the modes (\ref{eq:phikappa}) in the Fourier transform of the general solution (\ref{eq:Psi2}) recovers the left- and right-moving expression (\ref{eq:Fgen}), with $F_\pm(x) = \beta_\pm(-x)$. In particular, this means that the Gaussian wavepacket coefficients (\ref{eq:betag}) lead to the full wavefunction
\be\label{eq:psikappa}
\hat \Psi_{\pm, \Delta,\vep_o,k_o}(\kappa,k) = \frac{\bar c}{\sqrt{1+\k^2}} \,  e^{i \vep_o (k + k_o \mp \sinh^{-1}\k)} e^{- (k + k_o \mp  \sinh^{-1}\k)^2 \Delta^2/2} \,.
\ee
Recall that $\bar c$ was a normalization constant. Choosing the $\hat \Psi_+$ solution, the wavefunction (\ref{eq:psikappa}) explicitly represents a wavepacket peaked on the classical solutions $\kappa = \frac{1}{4} \pi_v = \frac{1}{4} \pa S/\pa v = \sinh[k+k_o]$, with no integrals remaining. This representation is therefore easy to use in computations of expectation values, as we now demonstrate. In the future interior, the momentum $\kappa$ increases monotonically from $-\infty$ at the horizon to $+\infty$ at the singularity, passing through zero at the maximal volume slice when $k = - k_o$.

The conserved norm in this description requires a moment's thought. The most obvious conserved quantity following from (\ref{eq:wdw3}) that distinguishes between right- and left-moving modes is the momentum of the wave. However, using this quantity would lead to a norm involving derivatives of the function $F$. In contrast, the norm we discussed previously in (\ref{eq:knorm}) is given, using the aforementioned fact that $F(x) = \beta(-x)$, by
\be\label{eq:nF}
|\hat \Psi_+|^2_\kappa = \int \frac{dk}{2\pi} |F_+(k)|^2 \,.
\ee
This expression is manifestly independent $\kappa$ and hence conserved. We can equivalently write (\ref{eq:nF}) as
\be
|\hat \Psi_+|^2_\kappa = \int \frac{dk}{2\pi} (1 + \kappa^2) |\hat \Psi_+(\kappa,k)|^2 \,.
\ee
This is the form we will use for calculating expectation values.
It should be noted that the norm we have constructed here is only defined on right-moving solutions to the wave equation. Therefore, these need to be selected a priori as the physical states. Previously we wrote down a norm that was conserved on all states but only positive on the physical ones.

The following expectation values are now immediate on the wavepacket (\ref{eq:psikappa})
\be
\langle k \rangle_{\kappa;\Delta,\vep_o,k_o} = - k_o + \sinh^{-1} \kappa \,, \qquad \langle \pi_k \rangle_{\kappa;\Delta,\vep_o,k_o} = \vep_o \,.
\ee
The expectation value of $k$ is seen to precisely obey the classical equation of motion without any corrections (in contrast to e.g.~(\ref{eq:kvd}) in the volume norm). The expectation value of $\pi_k$ is again conserved and equal to the value obtained with the other norms. The variances are also simple and given exactly by
\be
\text{var}(k)_{\kappa;\Delta,\vep_o,k_o} = \frac{1}{2 \Delta^2} \,, \qquad \text{var}(\pi_k)_{\kappa;\Delta,\vep_o,k_o} = \frac{\Delta^2}{2} \,.
\ee
Finally the `Hamiltonian' for the $\kappa$ clock, which from (\ref{eq:kap}) is just the negative of volume $\pi_\k = - 4 v$, has expectation value
\begin{align}
\langle \pi_\kappa \rangle_{\kappa;\Delta,\vep_o,k_o} & = - i \int \frac{dk}{2\pi} \sqrt{1+\k^2} \hat \Psi_{+, \Delta,\vep_o,k_o}^*(\k,k) \pa_\k \left(\sqrt{1+\k^2} \hat \Psi_{+, \Delta,\vep_o,k_o}(\k,k)\right)
\\
& = \frac{-\vep_o}{\sqrt{1 + \k^2}} \,.\label{eq:pikap}
\end{align}
The Wheeler-DeWitt equation (\ref{eq:wdw3}) depends explicitly on $\kappa$ and hence the Hamiltonian $\pi_\kappa$ is time-dependent. 
Equation (\ref{eq:pikap}) recovers the classical relation between the volume and extrinsic curvature -- see the discussion below (\ref{eq:pivv}).

\subsection{Metric component clocks}
\label{sec:metric}

Even within the two dimensional minisuperspace that is being considered, there are many other possible clocks involving combinations of $v$ (or $\kappa$) and $k$. We have focused on $v$ and $k$ in the above partly because the calculations are fairly straightforward. In this section we will consider a further natural choice of coordinates, the metric components
\be\label{eq:met}
g_{tt} = v^{2/3} e^{4 k/3} \,, \qquad g_{xx} = v^{2/3} e^{-2k/3} \,.
\ee
These coordinates will be useful shortly when we consider the extension to the exterior in \S\ref{sec:exterior}, which will amount to allowing $g_{tt}$ to be negative. 

One general aspect that may be important, especially once higher dimensional minisuperspaces are considered, is the sign of the `time' direction with respect to the DeWitt metric on superspace \cite{Halliwell:1989myn}. This metric determines the derivative structure of the Wheeler-DeWitt equation. For our two dimensional minisuperspace the DeWitt metric $G$ is, with the action normalised as in (\ref{eq:act}),
\be
G_{ab} \pi^a \pi^b = \frac{3}{8} \left( \frac{\pi_k^2}{v} - v \pi_v^2 \right)\,.
\ee
The volume direction is timelike, perhaps more intuitively associated to a clock, while the anisotropy direction is spacelike. In terms of the metric components (\ref{eq:met}) the DeWitt metric becomes
\be
G_{ab} \pi^a \pi^b = \sqrt{g_{tt}} \left(\frac{g_{tt}}{2 g_{xx}} (\pi_{tt})^2 -  \pi_{tt} \pi_{xx} \right) \,.
\ee
And the Wheeler-DeWitt equation is, again with the canonical choice of operator ordering in which the differential operator that appears is the Laplacian of the inverse DeWitt metric,
\be\label{eq:wdw4}
\frac{\pa}{\pa g_{tt}} \left(\frac{g_{tt}}{2 g_{xx}} \frac{\pa \Psi}{\pa g_{tt}} - \frac{\pa \Psi}{\pa g_{xx}}\right) + 6 g_{xx} \Psi = 0 \,.
\ee
The general solution to this equation can be written as
\be\label{eq:Psi3}
\Psi_\pm[g_{tt},g_{xx}] = \int \frac{d\bar k}{2\pi} \beta_\pm(\bar k) \exp\left[\pm 2 i \sqrt{g_{xx}} \left(e^{\bar k} g_{tt} - e^{-\bar k} g_{xx} \right) \right] \,.
\ee
The expression (\ref{eq:Psi3}) is, of course, exactly the same as (\ref{eq:Psi2}), now given in terms of the metric components instead of $v$ and $k$.

It will be useful to consider $g_{tt}$ as the relational time coordinate. Note that $g_{tt}$ is a spacelike direction under the inverse DeWitt metric. Within our two dimensional minisuperspace there is no intrinsic difference between spacelike and timelike directions. Further work is needed to establish whether spacelike `clocks' are admissible more generally. The conserved norm associated to the $g_{tt}$ clock from (\ref{eq:wdw4}) is
\be
|\Psi|^2_{g_{tt}} = \frac{-i}{2} \int dg_{xx} \left[\Psi^* \left(\frac{g_{tt}}{g_{xx}} \frac{\pa}{\pa g_{tt}} - \frac{\pa}{\pa g_{xx}}\right) \Psi - \Psi \left(\frac{g_{tt}}{g_{xx}} \frac{\pa}{\pa g_{tt}} - \frac{\pa}{\pa g_{xx}} \right) \Psi^* \right] \,.
\ee
This norm can be verified to vanish under $\pa/{\pa g_{tt}}$, assuming suitable vanishing of the wavefunction at large $g_{xx}$. On states of the form (\ref{eq:Psi3}) the norm becomes the same as (\ref{eq:knorm}) previously:
\be\label{eq:ngtt}
|\Psi_+|^2_{g_{tt}} = \int \frac{d\bar k}{2\pi} |\beta_+(\bar k)|^2 \,.
\ee
We may also calculate, for example, the expectation value of the corresponding `Hamiltonian'
\begin{align}
\langle \pi_{g_{tt}} \rangle_{g_{tt}} & = \frac{-1}{2} \int dg_{xx} \left[\Psi^* \frac{\pa}{\pa g_{tt}}\left(\frac{g_{tt}}{g_{xx}} \frac{\pa}{\pa g_{tt}} - \frac{\pa}{\pa g_{xx}}\right) \Psi - \frac{\pa \Psi}{\pa g_{tt}} \left(\frac{g_{tt}}{g_{xx}} \frac{\pa}{\pa g_{tt}} - \frac{\pa}{\pa g_{xx}} \right) \Psi^* \right] \\
 & = \int dg_{xx} \left[ \frac{g_{tt}}{2g_{xx}} \left|\frac{\pa \Psi}{\pa g_{tt}} \right|^2 + 6 g_{xx} |\Psi|^2 \right] \,.
\end{align}
These integrals are not as easy to evaluate at those in the $k$ and $v$ basis, performed above. However, at leading order on the Gaussian wavepacket the expectation values will obey the classical equations of motion. The Hamilton-Jacobi function is the exponent in (\ref{eq:Psi3})
\be\label{eq:Sfull}
S(g_{tt},g_{xx};k_o) =  2 \sqrt{g_{xx}} \left(e^{k_o} g_{tt} - e^{-k_o} g_{xx} \right) \,.
\ee
And therefore classically
\be
\pi_{g_{tt}} = 2 e^{k_o} \sqrt{g_{xx}} \,,
\ee
where $g_{xx}$ is classically related to $g_{tt}$ via
\be\label{eq:ttxx}
\vep_o = \pa_{k_o} S = 2 \sqrt{g_{xx}} \left(e^{k_o} g_{tt} + e^{-k_o} g_{xx} \right) \,.
\ee
We will discuss the quantum variance about this relation in \S\ref{sec:sing} below.

The $g_{xx}$ clock is a little more subtle because constant $g_{xx}$ slices are null under the inverse DeWitt metric. However, the conserved norm may still be defined via a limiting sequence of spacelike slices. One obtains the conserved norm
\be
|\Psi|^2_{g_{xx}} = \frac{-i}{2} \int dg_{tt} \left(\Psi^* \frac{\pa \Psi}{\pa g_{tt}} - \Psi \frac{\pa \Psi^*}{\pa g_{tt}} \right) \,.
\ee
It is easily verified that this vanishes under $\pa/\pa g_{xx}$, as always with assumptions about falloff at large $g_{tt}$. On states of the form (\ref{eq:Psi3}), this norm is equal to the previous expression (\ref{eq:ngtt}).

\section{The exterior and holographic renormalization}
\label{sec:exterior}

The solution (\ref{eq:Sfull}) to the classical Hamilton-Jacobi equation is valid inside or outside of the horizon. The two regions are distinguished by the sign of $g_{tt}$. In the exterior, the Hamilton-Jacobi equation is related to a radial, rather than timelike, slicing of spacetime. It has long been appreciated that in a holographic context, the function $S$ controls the holographic renormalization group flow \cite{deBoer:1999tgo, Heemskerk:2010hk, Faulkner:2010jy}. The arguments of $S$ are couplings of the dual quantum field theory and the momenta conjugate to these arguments are field theory expectation values of the corresponding dual operators.

Concretely, the basic holographic relation gives the quantum field theory (QFT) Lorentzian partition function as
\be\label{eq:Z}
Z_\text{QFT}[\gamma] = \int {\mathcal D} g e^{i S_\text{act}[g] + i S_\text{ct}[\gamma]} \,,
\ee
where the path integral over bulk metrics $g$ is restricted to those that are asymptotically AdS with conformal boundary metric $\gamma$. Note that within our minisuperspace we will have $\gamma_{tt} = g_{tt}, \gamma_{xx} = \gamma_{yy} = g_{xx}$, all constant in the boundary directions. There is no need to explicitly remove a conformal factor from the metric, as is often done. With $g_{tt}$ as the `clock', the conformal boundary is at $g_{tt} \to -\infty$, and conformal invariance will automatically constrain observables in this limit. However,
we need not take $g_{tt}$ all the way to infinity. Instead, we are interested in the holographic renormalization group flow that evolves as a function of $g_{tt}$. In (\ref{eq:Z}) we have included a boundary counterterm action \cite{Balasubramanian:1999re, Emparan:1999pm}
\be
S_\text{ct} = 4 \int d^3x \sqrt{-\gamma} = 4 \sqrt{-g_{tt}} g_{xx} \,.
\ee
The boundary metric is flat and hence no boundary curvature term is needed. We have normalized the volume of the boundary coordinates by setting $\int d^3x = 1$. In fact, we had implicitly already done this in our discussion of the Wheeler-DeWitt wavefunction above, where this normalization factor would otherwise appear in the conjugate momenta. Keeping the range of the boundary coordinates fixed is important for $g_{tt}$ and $g_{xx}$ to have an unambiguous meaning.

It will be instructive to quickly recast a standard holographic calculation in the Hamilton-Jacobi language. At leading order in the classical limit, the partition function (\ref{eq:Z}) is evaluated on the classical solution that tends to the boundary data $\gamma$. The action $S_\text{act}$ as a function of boundary data is precisely the Hamilton-Jacobi function (\ref{eq:Sfull}). Thus we obtain in the classical limit, including the counterterms in (\ref{eq:Z}),
\be\label{eq:logZ}
\log Z_\text{QFT}[g_{tt},g_{xx}; k_o] = - 2 i e^{-k_o}\sqrt{g_{xx}} \left(e^{k_o}\sqrt{-g_{tt}} - \sqrt{g_{xx}}\right)^2 \,.
\ee
Note that the partition function depends on the parameter $k_o$ of the classical solution, in addition to the boundary data. We will see shortly that this will allow the partition function to be computed in different ensembles. The energy density of the dual field theory is given by the momentum conjugate to $g_{tt}$,
\be\label{eq:vev}
\sqrt{-\gamma} \langle T^{t}{}_t \rangle_\text{QFT} = - 2 i \gamma_{tt} \frac{\pa \log Z_\text{QFT}}{\pa \gamma_{tt}} =
4 \sqrt{-g_{tt} g_{xx}} \left(\sqrt{g_{xx}} - e^{k_o} \sqrt{-g_{tt}}\right) \,.
\ee
We would like to evaluate this expectation value in the conformal field theory (CFT) limit $g_{tt} \to - \infty$, wherein the boundary is taken to infinity. To do this, we may use the classical relation between $g_{xx}$ and $g_{tt}$ given in (\ref{eq:ttxx}). 
In addition to $k_o$, this classical relation introduces a further independent constant $\vep_o$. At this point the value of $\vep_o$ is arbitrary; we will see shortly how it is determined from a given distribution of values of $k_o$.
As $g_{tt} \to - \infty$ we have from (\ref{eq:ttxx}) that
\be\label{eq:nearb}
g_{xx} = - e^{2 k_o} g_{tt} + \frac{\vep_o}{2 \sqrt{-g_{tt}}} + \cdots  \,.
\ee
The leading near-boundary behavior $g_{xx} \sim g_{tt}$
is controlled by the asymptotic CFT scaling. Using (\ref{eq:nearb}) in (\ref{eq:vev}) we obtain
\be\label{eq:EE}
\lim_{g_{tt} \to - \infty} \sqrt{-\gamma}  \langle T^{t}{}_t \rangle_\text{QFT} = \vep_o \,.
\ee
Thus we recover the fact that $\vep_o$ is the energy density of the black hole, as seen from the exterior boundary at infinity.

Another useful way to understand (\ref{eq:EE}) is to note that the partition function (\ref{eq:logZ}) is invariant under
\be\label{eq:shift}
\log Z_\text{QFT}[g_{tt} e^{- 4 \bar k/3}, g_{xx} e^{2 \bar k/3}; k_o + \bar k] = \log Z_\text{QFT}[g_{tt},g_{xx}; k_o] \,.
\ee
This invariance can be seen explicitly in (\ref{eq:logZ}) but follows more generally from the definitions (\ref{eq:met}) and the shift symmetry in $k$. Recall that $\vep$ was first introduced in (\ref{eq:diffS}) as the conjugate variable to $k_o$. Differentiating (\ref{eq:shift}) with respect to $\bar k$ and then setting $\bar k = 0$ therefore gives
\begin{align}
\langle \vep \rangle_\text{QFT} & = - i \frac{\pa \log Z_\text{QFT}}{\pa k_o} = \frac{- 2 i }{3} \left(2 g_{tt} \frac{\pa \log Z_\text{QFT}}{\pa g_{tt}} - g_{xx}\frac{\pa \log Z_\text{QFT}}{\pa g_{xx}} \right) \label{eq:aa} \\
& = \frac{2 \sqrt{-\gamma}}{3}\left(\langle T^{t}{}_t \rangle_\text{QFT} - \langle T^{x}{}_x \rangle_\text{QFT} \right) \,. \label{eq:bb}
\end{align}
In the final equality we used the fact that $\pa \log Z_\text{QFT}/\pa g_{xx}$ is two times $\langle T^{x}{}_x \rangle_\text{QFT}$, because in computing $\log Z_\text{QFT}$ we set $g_{xx} = g_{yy}$ while these must be kept distinct when taking the derivatives to obtain $\langle T^{x}{}_x \rangle_\text{QFT}$. In the 
CFT limit of $g_{tt} \to - \infty$ we may then use the vanishing of the trace of the energy-momentum tensor: $\langle T^{t}{}_t \rangle_\text{QFT} + 2 \langle T^{x}{}_x \rangle_\text{QFT} = 0$ (our background is flat, there is no anomaly). Using this relation we recover (\ref{eq:EE}) from the final line of (\ref{eq:bb}).

From (\ref{eq:aa}) and (\ref{eq:bb}) we learn that while $k_o$ is not boundary time in general, because it generates a transformation of both space and time, it is equivalent to time in the CFT limit where the boundary is taken to infinity. In particular, we can write the partition function of the asymptotic boundary theory as
\be\label{eq:btime}
Z_\text{QFT}[g_{tt},g_{xx}; k_o] = \Tr \left( e^{i k_o H_\text{QFT}[g_{tt},g_{xx}]} \right) \,.
\ee
Here $H_\text{QFT}[g_{tt},g_{xx}]$ is the dual field theory Hamiltonian in a given background metric $g_{tt}$ and $g_{xx}$ (both of which are constant within our minisuperspace). This relation will be useful shortly. Away from the CFT limit one should replace the Hamiltonian in (\ref{eq:btime}) with the combination of $T^{t}{}_t$ and $T^x{}_{x}$ appearing in (\ref{eq:bb}).

Beyond the classical limit the partition function (\ref{eq:Z}), with the boundary counterterms subtracted out, obeys the same Wheeler-DeWitt equation that we have been discussing throughout. This follows from the usual connection between path integrals and wavefunctions (e.g.~\cite{Hartle:1983ai}). We do not usually think of the boundary partition function as a state because the slicing in the bulk is radial rather than in time. However, if we consider $g_{tt}$ as a clock then formally the boundary partition function will arise as the $g_{tt} \to - \infty$ limit of the interior wavefunction $\Psi[g_{tt},g_{xx}]$. Alternatively, we may think of the boundary partition function as setting boundary conditions on the interior wavefunction. This relation is shown in Fig.~\ref{fig:penrose}. There may be fruitful connections here to discussions of $T^2$ deformations of the boundary theory, e.g.~\cite{McGough:2016lol, Hartman:2018tkw}. The objective of those works is to identify field-theoretic deformations that correspond to moving the partition function into the bulk.

Within the minisuperspace approximation, we may therefore represent the semiclassical partition function as
\be\label{eq:Z2}
Z_\text{QFT}[g_{tt},g_{xx}; \beta] = e^{4 i \sqrt{-g_{tt}} g_{xx}} \Psi_+[g_{tt},g_{xx}; \beta] \,,
\ee
where $\Psi_+$ is the general solution already given in (\ref{eq:Psi3}). This solution depends on an arbitrary function $\beta_+(\bar k)$. What is the boundary interpretation of this function? We have seen in (\ref{eq:btime}) that $\bar k$ generates boundary time translations. Therefore from (\ref{eq:Z2}) and the wavefunction (\ref{eq:Psi2})
\be
Z_\text{QFT}[g_{tt},g_{xx}; \beta] = \int \frac{d\bar k}{2\pi} \beta_+(\bar k) \Tr \left( e^{i \bar k H_\text{QFT}[g_{tt},g_{xx}]} \right) \,.
\ee
In particular, consider the Gaussian wavepacket (\ref{eq:betag}). This leads to
\begin{align}
Z_\text{QFT}[g_{tt},g_{xx}; k_o, \vep_o, \Delta] & = \bar c \int \frac{d\bar k}{2\pi} e^{- i \vep_o (\bar k - k_o)} e^{- (\bar k - k_o)^2 \Delta^2/2} \Tr \left( e^{i \bar k H_\text{QFT}[g_{tt},g_{xx}]} \right) \\
& = \frac{c}{\sqrt{2 \pi}} \text{Tr} \left[ e^{- (H_\text{QFT}[g_{tt},g_{xx}] - \vep_o)^2/(2 \Delta^2)} e^{i k_o H_\text{QFT}[g_{tt},g_{xx}]} \right] \,. \label{eq:Gau}
\end{align}
The final expression (\ref{eq:Gau}) is the partition function of a microcanonical energy window.

We may verify (\ref{eq:Gau}) explicitly in the leading order semiclassical limit. We start from the wavefunction in (\ref{eq:Z2}) with the Gaussian weighting from (\ref{eq:betag}), Fourier transform the Gaussian and then perform the $\bar k$ integral by stationary phase. Finally, considering the near-boundary CFT limit $g_{tt} \to - \infty$ leads to
\be
\lim_{g_{tt} \to - \infty}Z_\text{QFT}[g_{tt},g_{xx}; k_o, \vep_o, \Delta] \sim
\int d\vep \, \exp\left\{-\frac{(\vep - \vep_o)^2}{2 \D^2} +  i\vep \log \frac{e^{k_o} \sqrt{-g_{tt}}}{\sqrt{g_{xx}}} + \frac{i \vep^2}{8 g_{xx} \sqrt{-g_{tt}}}\right\} \,.
\ee
Here $\sim$ means to leading classical order without keeping track of the prefactor. We can note that the $\vep^2$ in the final term is the correct scaling for the density of states in the dual 2+1 dimensional conformal field theory. The second term comes from the smearing of the `time' $\bar k$ about the mean value $k_o$. This effect has also changed the sign of the final term.

Equations (\ref{eq:Z2}) and (\ref{eq:Gau}) show how an interior Wheeler-DeWitt wavefunction is directly encoded in a boundary partition function. The boundary field theory limit is $g_{tt} \to - \infty$, while the near-singularity regime is $g_{tt} \to + \infty$. We can note that the underlying Hamilton-Jacobi function (\ref{eq:Sfull}) is a linear function of $g_{tt}$, so that the two limits are connected in a straightforward way. Nonetheless, the limits reveal different physics. We saw the near-boundary CFT scaling emerge in (\ref{eq:nearb}). In contrast, as $g_{tt} \to + \infty$ we have from the classical solution (\ref{eq:ttxx}) that
\be\label{eq:nearsing}
g_{xx} = \frac{\vep_o^2 e^{-2 k_o}}{4 g_{tt}^2} + \cdots \,,
\ee
showing the near-singularity collapse of two space directions. The scaling here is controlled by the interior Kasner exponent \cite{Frenkel:2020ysx}.

One may hope that this connection between the interior and the boundary will provide a firm foundation for understanding the physics of `timeless' cosmological Wheeler-DeWitt states. Conversely, this connection shows how basic boundary observables can contain signatures of the interior, including the spacelike singularity. We now elaborate on this point.

\section{Behavior near the singularity}
\label{sec:sing}

Spacelike singularities result in the `end of time' and therefore raise challenging interpretational questions. One might think that the end of time is best understood in a framework in which there is no time to begin with, such as the Wheeler-DeWitt equation. Once time is understood relationally, the end of time is simply a limit to the relations that can exist.

The computations of expectation values performed in earlier sections, with various different notions of relational time, did not show any breakdown of minisuperspace classicality as the singularity was approached. In particular, the quantum variance associated with the wavepacket remained small as the singularity was approached. To further emphasize this point we can consider the expectation value of a diverging curvature. The Weyl curvature squared may be expressed within our minisuperspace as
\be\label{eq:W2}
W_{abcd} W^{abcd} = \frac{3 \pi_k^2}{16 v^4} \left(\pi_k + v \pi_v \right)^2 \,.
\ee
To obtain this expression we have eliminated a time derivative of a momentum using the Hamiltonian equations of motion $\pa_r \pi_k = - \pa_k H = 0$. Here $H$ is the Hamiltonian following from (\ref{eq:Lag}). This step is necessary because the spacetime coordinate $r$ does not appear in the timeless quantum theory. In any event, the elimination is especially simple in this case because, as we have already noted above, $\pi_k$ is conserved.

We may compute the expectation value and variance of (\ref{eq:W2}) using, for example, the anisotropy time of \S\ref{sec:aniso}. Volume time is not convenient because the volume tends to zero at both the singularity and the horizon. York time is also inconvenient here because of the factor of $1/v^4$ in (\ref{eq:W2}), which is an inverse factor of the York momentum. The expectation value (in the limit (\ref{eq:range})) is the classical behavior
\be
\left\langle \frac{3 \pi_k^2}{16 v^4} \left(\pi_k + v \pi_v \right)^2 \right\rangle_{k;\Delta,\vep_o,k_o} = 12 \left( 1 + e^{2 (k_o + k)} \right)^2 \,.
\ee
As expected the curvature diverges at the singularity where $k \to + \infty$ but is regular at the horizon where $k \to - \infty$. The variance is somewhat tedious to compute but is given, to leading order in the limit (\ref{eq:range}), by
\be
\text{var} \left[ \frac{3 \pi_k^2}{16 v^4} \left(\pi_k + v \pi_v \right)^2 \right]_{k;\Delta,\vep_o,k_o} = \frac{1152}{\Delta^2} e^{4 (k_o+k)} \left( 1 + e^{2 (k_o + k)} \right)^2 \,.
\ee
In this computation we chose the manifestly Hermitian operator ordering $\frac{1}{2}(v \pi_v + \pi_v v)$ for the $v \pi_v$ term.
Thus we see that, consistently with all of our earlier results, the variance remains uniformly small compared to the square of the expectation value when $\Delta \gg 1$.

As a final example, we can consider the variance in $g_{xx}$ as this spatial direction collapses towards the singularity. With $g_{tt}$ as the clock, the classical behavior towards the singularity has already been given in (\ref{eq:nearsing}). Using the expressions in \S\ref{sec:metric} the variance is found to be, as $g_{tt} \to + \infty$,
\be
\text{var} \left[ g_{xx} \right]_{g_{tt};\Delta,\vep_o,k_o} = \frac{2 \Delta^2}{\vep_o^2} \left\langle g_{xx} \right\rangle_{g_{tt};\Delta,\vep_o,k_o}^2 \,.
\ee
The variance is therefore again uniformly small compared to the expectation value in the limit (\ref{eq:range}). Minisuperspace quantum fluctuations do not keep this dimension from collapsing towards the singularity.

We wish to emphasize two related points from the discussion above. The first is that the Gaussian wavepacket is able to probe the singularity. We do not find the phenomenon, reported in \cite{Perry:2021mch, Perry:2021udd}, of the wavefunction vanishing towards the singularity. This is promising from the point of view of identifying signatures of the singularity in the boundary partition functions discussed in \S\ref{sec:exterior}. Secondly, the quantum uncertainty due to the wavepacket is not enough to `resolve' the singularity, at least within the minisuperspace description. As we have mentioned at the start of this section, within a timeless quantum state there is not necessarily anything pathological about the end of time. However, it is possible that the spread of the wavepacket will play a more prominent role once inhomogeneous fluctuations are incorporated. Furthermore, as commented in \S\ref{sec:future} below, it would be natural for microscopic degrees of freedom --- beyond the semiclassical metric --- to become relevant in the wavefunction as the singularity is approached.

\section{Discussion}
\label{sec:future}

This work has been an exploration of the Wheeler-DeWitt equation in black hole interiors. We have limited ourselves to a minisuperspace description of one of the best-understood of all black holes --- planar AdS-Schwarzschild. The simplicity of the description, including an explicit solution to the Wheeler-DeWitt equation, has allowed us to focus on conceptual issues such as the emergence of relational notions of time and the connection between the interior and the exterior. However, to properly confront questions regarding the black hole singularity it will be essential to go beyond minisuperspace.

It is helpful to contrast the situation here with the Wheeler-DeWitt equation as it arises in de Sitter space in e.g.~\cite{Hartle:1983ai, Maldacena:2002vr, Anninos:2012ft, Pimentel:2013gza}. A large volume, classical and homogeneous universe is a reasonable starting point for understanding the late time wavefunction of de Sitter space. However, the approach to interior singularities is expected to be highly inhomogeneous \cite{Belinsky:1970ew}. Because homogeneous interiors, such as the ones we have been considering, are unstable towards developing inhomogeneities, they likely do not define a useful background geometry on top of which to construct a holographic correspondence for late interior times. However, it may be interesting to incorporate quantum effects as inhomogeneities become important in the interior by solving the fully inhomogeneous Wheeler-DeWitt equation. Furthermore, as curvatures grow towards the singularity microscopic `stringy' degrees of freedom can be expected to become relevant. One way to characterize the resolution of spacelike singularities would be to write down a Wheeler-DeWitt wavefunction in a fully microscopic theory. It may be interesting to revisit ideas involving tachyon condensation or matrix degrees of freedom near cosmological singularities --- see \cite{McAllister:2007bg} for an overview with references --- from a `timeless' wavefunction perspective.

Even within the simplified minisuperspace description there remain interesting issues to explore. In particular, Cauchy horizons raise additional conceptual challenges because the interior is no longer contained within a single causal patch. One manifestation of this fact is that the $g_{tt}$ relational clock is no longer single-valued in the interior because it vanishes at both the outer and inner horizons. On the other hand, it may be that the causal patch between the two horizons corresponds to a self-contained timeless Wheeler-DeWitt state. From this perspective, Cauchy horizons may not be a problem that needs to be solved. To examine these questions, it may be interesting to study minisuperspace Wheeler-DeWitt states in Einstein-Maxwell interiors and in other simple holographic models with and without classical Cauchy horizons \cite{Hartnoll:2020rwq, Hartnoll:2020fhc}.

One theme of this work has been that the Hamilton-Jacobi formulation of dynamics is especially well-suited to extension through horizons. While the boundary time coordinate becomes complex in the black hole interior, cf.~\cite{Motl:2003cd, Fidkowski:2003nf, Hartman:2013qma, Grinberg:2020fdj}, the Hamilton-Jacobi formulation makes no reference to time, or any other coordinate, but instead expresses relations between physical variables. These variables, such as $g_{tt}$, can be relationally evolved through the horizon without difficulty --- at least in the minisuperspace description that we have considered. This may suggest a way to search for interior dynamics in the dual field theory. Instead of analytically continuing Green's functions of time or energy, one should compute suitable energy window Lorentzian partition functions of the field theory and consider their behavior at positive values of $g_{tt}$. If this works, then large $N$ field theories may contain within themselves, fairly explicitly, the rich inhomogeneous dynamics of black hole interiors and the resolution of interior singularities.

Another theme we have emphasized is that the exterior provides a well-defined anchor for interior Wheeler-DeWitt wavefunctions. This can be contrasted with controversies regarding the appropriate boundary conditions in quantum cosmology. Different boundary partition functions prepare different interior wavefunctions via e.g.~(\ref{eq:Z2}). It may be interesting to ask how the interior wavefunction contains physics such as the black hole butterfly effect \cite{Shenker:2013pqa}, that arises when the exterior involves sources on both sides of the thermofield double.

It is interesting that quantities closely related to boundary time and energy --- $k_o$ and $\vep_o$ --- appeared as constants of integration in the Hamilton-Jacobi description, and correspondingly as parameters in the Wheeler-DeWitt wavepackets that we constructed. In the quantum mechanical description these parameters need not necessarily be real. In particular, it may be interesting to consider imaginary $k_o$ as this will be related to a canonical, fixed temperature, boundary partition function via (\ref{eq:btime}).

Finally, we have recalled how relational clocks are useful because the structure of the Wheeler-DeWitt equation ensures the existence of an associated conserved probability. We then recalled how positivity of probabilities requires the Hilbert space to be built out of `half' the modes, analogous to the positive energy modes in quantum field theory. If modes of opposite sign are respectively associated to the future and past interiors of the black hole then this may seem to preclude the possibility of a `bounce' in which a classical future interior connects to a classical past interior through a quantum mechanical regime. Such bounces have been widely discussed in `loop quantum cosmology', recently reviewed in \cite{Ashtekar:2022oyq}. It may be interesting to determine precisely what kinds of quantum bounces are possible within sign-constrained superspace. Furthermore, related to these points, while we have taken the Klein-Gordon-like norm of the Wheeler-DeWitt equation seriously, and built first-quantized theories around that, it may be possible to consider a more second-quantized formalism that would allow phenomena such as pair production of interior universes.

\section*{Acknowledgements}

I am especially grateful to Mahdi Godazgar for suggesting the Wheeler-DeWitt equation as a probe of the interior, and to Raghu Mahajan for very helpful comments on a first draft of this paper. It is furthermore a great pleasure to acknowledge helpful discussions with Dio Anninos, Alex Belin, Frederik Denef, Ronak Soni and Aron Wall. This work was partially supported by Simons Investigator award \#620869 and by STFC consolidated grant ST/T000694/1.

\providecommand{\href}[2]{#2}\begingroup\raggedright\endgroup

\end{document}